\documentclass{article}
\usepackage{amssymb,amsmath}
\usepackage{epsfig}
\usepackage{pdflscape}
\usepackage{subfigure}
\usepackage{color}
\usepackage{lscape}
\usepackage{bbm}
\usepackage{bm}
\usepackage{mathrsfs}
\usepackage{amsfonts}
\usepackage{graphics}
\usepackage{indentfirst}
\usepackage[square,numbers,sort&compress]{natbib}
\usepackage{multicol}

\newtheorem{rem}{Remark}[section]
\newcommand{\br}{\begin{rem}}
\newcommand{\er}{\end{rem}}
\newtheorem{ex}[rem]{Example}
\newcommand{\bex}{\begin{ex}}
\newcommand{\eex}{\end{ex}}
\newtheorem{Def}[rem]{Definition}
\newcommand{\bd}{\begin{Def}}
\newcommand{\ed}{\end{Def}}
\newtheorem{theorem}[rem]{Theorem}
\newcommand{\bt}{\begin{theorem}}
\newcommand{\et}{\end{theorem}}
\newtheorem{Prop}[rem]{Proposition}
\newcommand{\bp}{\begin{Prop}}
\newcommand{\ep}{\end{Prop}}
\newtheorem{lemma}[rem]{Lemma}
\newcommand{\bl}{\begin{lemma}}
\newcommand{\el}{\end{lemma}}
\newcommand{\be}{\begin{equation}}
\newcommand{\ee}{\end{equation}}
\newcommand{\bea}{\begin{eqnarray}}
\newcommand{\eea}{\end{eqnarray}}
\newcommand{\pa}{\partial}
\newcommand{\nn}{\nonumber}
\newcommand{\adots}{\mathinner{\mkern2mu\raise1pt\hbox{.}\mkern2mu
\raise4pt\hbox{.}\mkern2mu\raise7pt\hbox{.}\mkern1mu}}

\title{The Role of Commuting Operators\\ in Quantum Superintegrable Systems}

\author{Allan P. Fordy, School of Mathematics,\\
University of Leeds, Leeds LS2 9JT, UK.\\ ~~E-mail: a.p.fordy@leeds.ac.uk}

\begin{document}

\maketitle

\begin{abstract}
We discuss the role of commuting operators for quantum superintegrable systems, showing how they are used to build eigenfunctions.  These ideas are illustrated in the context of resonant harmonic oscillators, the Krall-Sheffer operators, with polynomial eigenfunctions, and the Calogero-Moser system with additional harmonic potential. The construction is purely algebraic, avoiding the use of separation of variables and differential equation theory.
\end{abstract}




\section{Introduction}

The notions of complete and superintegrability have their origins in classical mechanics.

A Hamiltonian system of $n$ degrees of freedom, Hamiltonian $H$, is said to be {\em completely integrable in the Liouville sense} if we have $n$ independent functions $I_n$, which are {\em in involution} (mutually Poisson commuting), with $H$ being a function of these and typically just one of them. Whilst $n$ is the maximal number of independent functions which can be {\em in involution}, it is possible to have further integrals of the Hamiltonian $H$, which necessarily generate a non-Abelian algebra of integrals of $H$.  The maximal number of additional {\em independent} integrals is $n-1$, since the ``level surface'' of $2n-1$ integrals (meaning the intersection of individual level surfaces) is just the (unparameterised) integral curve.  Well known elementary examples are the isotropic harmonic oscillator, the Kepler system and the Calogero-Moser system.

The idea can be extended to {\em quantum integrable systems}, with first integrals replaced by commuting differential operators.  If we are truly interested in the system as a physical quantum system, then we would require our operators to be Hermitian.  However, the same ideas are applicable to the study of polynomial eigenfunctions of Laplace-Beltrami operators, so we will not impose this condition.  Whereas the goal in {\em classical} mechanics is to solve the equations of motion for the {\em trajectory}, the {\em quantum} case seeks the {\em spectrum} and corresponding {\em eigenfunctions}.  Superintegrability leads to degeneration of eigenvalues in a {\em direct way}, to be explained in this paper.

A commuting operator can share eigenfunctions, but can also {\em permute} the eigenfunctions of a degenerate eigenvalue of the Hamiltonian.  The best known example of this is the action of the rotation algebra on the eigenfunctions of a spherically symmetric Hamiltonian, such as that of the hydrogen atom.  This is, indeed, the simplest example, since the rotation algebra is represented by first order differential operators.  Another example is the resonant harmonic oscillator, which involves higher order differential operators, as described in Section \ref{Sec:ResHarmonic}.  This phenomenon is commonly seen in superintegrable quantum systems, where it is possible to use the additional commuting operators to {\em explicitly} build sequences of eigenfunctions \cite{f05-1,f06-1,f07-1}.

There is a large literature on the classification and analysis of superintegrable systems (see the review \cite{13-2}) and they naturally occur in many applications in physics (additional integrals being referred to as ``hidden symmetries'' \cite{14-2}).

In Section \ref{Sec:ResHarmonic} we briefly explain ideas in the familiar context of the {\em resonant harmonic oscillator}.  In Section \ref{Sec:KS2D} we discuss the Krall-Sheffer operators, which are deformations of the Laplace-Beltrami operators of associated (pseudo-)Riemannian manifolds.  The role of the isometries in building higher order commuting operators is emphasised.  The role of these operators in constructing a triangular array of polynomial eigenfunctions is explained.  In Section \ref{Sec:KS3D} we discuss the extension of Krall-Sheffer operators to a 3 dimensional domain.  Again the isometries play a key role in the building of commuting operators, which are then used to build a tetrahedral array of polynomial eigenfunctions.

We next consider the quantum Calogero-Moser system (in the 2 and 3 dimensional cases), with the addition of an external harmonic potential.  This is well known to be superintegrable and has a plethora of commuting operators.  By considering this as a deformation of the {\em isotropic harmonic oscillator}, we extend some of the latter's integrals to the Calogero-Moser case.  These operators are then used to build eigenfunctions, which are deformations of those of the isotropic harmonic oscillator.

\section{Resonant Harmonic Oscillators}\label{Sec:ResHarmonic}

Since two harmonic oscillators in Cartesian coordinates are just written as the sum of 1-dimensional oscillators, the system is clearly separable in both the classical and quantum cases.  This system can be explicitly solved by elementary methods.  When the oscillators are resonant, some additional features arise.  The classical orbits are periodic, forming Lissajous figures in the configuration space.  The quantum spectrum is degenerate, with multiple eigenfunctions corresponding to each eigenvalue.  Both these phenomena are a result of superintegrability (see \cite{f18-3} for further discussion).  Here we just discuss the quantum case.

We consider the eigenvalue problem
$$
L\psi\equiv \left(\pa_x^2+\pa_y^2-\omega_1^2 x^2-\omega_2^2 y^2\right) \psi = \lambda \psi,
$$
with ladder operators $A_x^\pm = \mp \pa_x + \omega_1 x,\, A_y^\pm = \mp \pa_y + \omega_2 y$, satisfying
$$  
[L,A_x^\pm]=\mp 2 \omega_1 A_x^\pm,\quad [L,A_y^\pm]=\mp 2 \omega_2 A_y^\pm,\quad [A_x^\pm,A_y^\pm]=0.
$$   
The ground state $\psi_{0,0}=e^{-\frac{1}{2}(\omega_1x^2+\omega_2y^2)}$ satisfies
$$
A_x^-\psi_{0,0}=A_y^-\psi_{0,0}=0,\quad L\psi_{0,0}= -(\omega_1+\omega_2)\psi_{0,0},
$$
and we use the raising operators to define an infinite triangular array of eigenfunctions:
$$
\psi_{i,j} = (A_x^+)^i (A_y^+)^j \psi_{0,0}, \quad\mbox{with eigenvalues}\;\; \lambda_{i,j}=-(2i+1)\omega_1-(2j+1)\omega_2.
$$
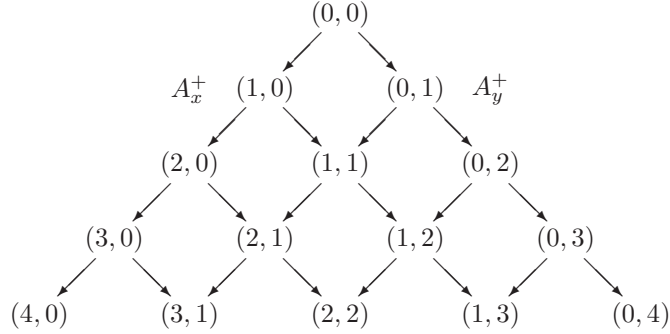
\begin{figure}[hbt]
\begin{center}
\caption{The action of $A_x^+$ and $A_y^+$ on the array $\psi_{(j,k)}$} \label{2dHarray}\vspace{6mm}
\unitlength=0.5mm
\begin{picture}(160,80)
\put(80,80){\makebox(0,0){$(0,0)$}}
\put(60,60){\makebox(0,0){$(1,0)$}}
\put(100,60){\makebox(0,0){$(0,1)$}}
\put(40,40){\makebox(0,0){$(2,0)$}}
\put(80,40){\makebox(0,0){{$(1,1)$}}}
\put(120,40){\makebox(0,0){$(0,2)$}}
\put(20,20){\makebox(0,0){$(3,0)$}}
\put(60,20){\makebox(0,0){$(2,1)$}}
\put(100,20){\makebox(0,0){$(1,2)$}}
\put(140,20){\makebox(0,0){$(0,3)$}}
\put(0,0){\makebox(0,0){$(4,0)$}}
\put(40,0){\makebox(0,0){{$(3,1)$}}}
\put(80,0){\makebox(0,0){$(2,2)$}}
\put(120,0){\makebox(0,0){{$(1,3)$}}}
\put(160,0){\makebox(0,0){$(0,4)$}}
\put(75,75){\vector(-1,-1){10}}
\put(85,75){\vector(1,-1){10}}\put(65,55){\vector(1,-1){10}}\put(95,55){\vector(-1,-1){10}}
\put(55,55){\vector(-1,-1){10}}\put(45,35){\vector(1,-1){10}}  \put(105,55){\vector(1,-1){10}}
\put(35,35){\vector(-1,-1){10}} \put(75,35){\vector(-1,-1){10}} \put(115,35){\vector(-1,-1){10}} \put(125,35){\vector(1,-1){10}}\put(25,15){\vector(1,-1){10}}\put(85,35){\vector(1,-1){10}}\put(65,15){\vector(1,-1){10}}\put(105,15){\vector(1,-1){10}}
\put(15,15){\vector(-1,-1){10}}\put(55,15){\vector(-1,-1){10}} \put(95,15){\vector(-1,-1){10}} \put(135,15){\vector(-1,-1){10}} \put(145,15){\vector(1,-1){10}}
\put(40,60){\makebox(0,0){$A_x^+$}}
\put(120,60){\makebox(0,0){$A_y^+$}}
\end{picture}
\end{center}
\end{figure}
Since the ladders in the $x-$direction commute with those in the $y-$direction, it doesn't matter which order we operate with $A_x^+$ and $A_y^+$.  Each order corresponds to a different path through the array depicted in Figure \ref{2dHarray}.  We can also move in the negative direction $A_x^-\psi_{j,k}=2 i \omega_1 \psi_{j-1,k},\, A_y^-\psi_{j,k}=2 i \omega_2 \psi_{j,k-1}$.

\medskip
{\em The eigenvalues $\lambda_{i,j}$ are distinct, if and only if $\omega_1$ and $\omega_2$ are {\em not} commensurate.}

\medskip
When $(\omega_1,\omega_2)=(m,n)$, we can build two commuting operators (see Jauch and Hill \cite{40-2}):
$$
[L,(A_x^\pm)^n] = \mp 2mn (A_x^\pm)^n,\quad [L,(A_y^\pm)^m] = \mp 2mn (A_y^\pm)^m\quad\Rightarrow\quad \left\{\begin{array}{l}
                                                                                                           \left[L, (A_y^+)^m (A_x^-)^n \right] = 0 ,\\[2mm]
                                                                                                                \left[L,(A_x^+)^n (A_y^-)^m\right] = 0.
                                                                                                              \end{array}\right.
$$
These play the role of connecting eigenfunctions with the {\em same} eigenvalue, so are directly related to {\em degeneracy}.

\paragraph{Isotropic Case: $(m,n)=(1,1)$.}  Here we have
$$
A_y^+A_x^- = x y+y\pa_x-x\pa_y-\pa_x\pa_y \quad\mbox{and}\quad  A_x^+A_y^- = x y-y\pa_x+x\pa_y-\pa_x\pa_y,
$$
taking us respectively right and left across a horizontal, with $i+j=\ell$, each with eigenvalue $\lambda_{i,j}=-2(\ell+1)$.

The more usual operators are just the even and odd combinations:
$$
\frac{1}{2} \left(A_y^+A_x^- + A_x^+A_y^-\right) = x y-\pa_x\pa_y,\quad   \frac{1}{2} \left(A_y^+A_x^- - A_x^+A_y^-\right) = y\pa_x-x\pa_y.
$$

This case (and its 3 dimensional version) arises in the context of the Calogero-Moser system of Sections \ref{Sec:2D-CM} and \ref{Sec:3D-CM}.

\paragraph{The Case: $(m,n)=(1,2)$.}  Here we have
\bea
A_y^+(A_x^-)^2 &=& 2 (x^2+1) y+4 x y\pa_x-(x^2+1)\pa_y+2 y \pa_x^2-2 x \pa_x\pa_y-\pa_x^2\pa_y,  \nn\\
(A_x^+)^2A_y^- &=& 2 (x^2-1) y-4 x y\pa_x+(x^2-1)\pa_y+2 y \pa_x^2-2 x \pa_x\pa_y+\pa_x^2\pa_y,  \nn
\eea
taking us respectively right and left in the direction of the line $i+2j=\ell$, each with eigenvalue $\lambda_{i,j}=-(2\ell+3)$.  For example, the first of these acts as $\psi_{i,j}\mapsto\psi_{i-2, j+1}$.

The even and odd parts are now respectively second and third order differential operators.

\subsection{Associated Hermite Polynomials}\label{Sec:ResHarmonicHermite}

As in the 1 dimensional case, we can set $\psi_{j,k} = \psi_{0,0} P_{j,k}(x,y)$, to obtain the 2-dimensional Hermite equation
\be\label{2dHermite}
\tilde L P\equiv \left(\pa_x^2+\pa_y^2-2\omega_1 x \pa_x-2\omega_2 y \pa_y-(\omega_1+\omega_2)\right) P  = \lambda P,
\ee
with the eigenfunctions being separated, with $P_{j,k}(x,y)=H_j(x)H_k(y)$, where $H_j,\, H_k$ are just the Hermite polynomials (but with coefficients depending upon $\omega_1$ or $\omega_2$).  Similarly the ladder operators reduce to the standard ones for Hermite polynomials: $\tilde A_x^+ = 2 \omega_1 x-\pa_x,\; \tilde A_x^- = \pa_x$ and so on.

These polynomials will play an important role in Sections \ref{Sec:2D-CM} and \ref{Sec:3D-CM}.

\section{Krall-Sheffer Operators and 2 Dimensional Polynomials}\label{Sec:KS2D}

Krall and Sheffer \cite{67-2} considered the class of 2 dimensional linear operators that could support polynomial solutions (a simple generalisation of Bochner's classification in 1 dimension):
\begin{subequations}
\be                               \label{admiss}
\begin{array}{l}
L \varphi:= (\alpha x^2 + d_1 x + e_1 y+f_1) \varphi_{xx} + (2\alpha x y + d_2 x + e_2 y+f_2) \varphi_{xy}\\[2mm]
 \hspace{2cm} + (\alpha y^2 +d_3 x + e_3 y+f_3) \varphi_{yy} + (\beta x +\kappa_1) \varphi_x + (\beta y + \kappa_2) \varphi_y.
 \end{array}
\ee
The degree $N$ polynomial eigenfunctions form an $N+1$ dimensional vector space, with a basis of {\em monic} polynomials:
\be  \label{pmn}  %
P_{m,n} = x^m y^n + \mbox{lower order terms} ,\quad\mbox{for}\;\;\; m+n=N ,
\ee  %
with eigenvalue
\be\label{lamN}
\lambda_{N}= \lambda_{m+n}=(m+n) ((m+n-1)\alpha+\beta).
\ee
\end{subequations}
The number of parameters in (\ref{admiss}) is reduced by making {\em affine transformations} on the $x-y$ space, giving $9$ canonical forms, given in Table \ref{Tab:KSops}.
\begin{table}[h]
\begin{center}
\caption{Krall-Sheffer Operators}\label{Tab:KSops}\vspace{3mm}
\begin{tabular}{|l|l|}
\hline
&  \\
Type I:  &  $L = (x^2-x) \pa_x^2+2 x y \pa_x\pa_y+ (y^2-y) \pa_y^2+ (\beta x+\kappa_1)\pa_x+ (\beta y+\kappa_2)\pa_y)$\\[2mm]
Type II: & $L = x^2 \pa_x^2+2 x y \pa_x\pa_y+ (y^2-y) \pa_y^2+ (\beta x+\kappa_1)\pa_x+ (\beta y+\kappa_2)\pa_y$ \\[2mm]
Type III:  & $L = x^2 \pa_x^2+2 x y \pa_x\pa_y+ (y^2+x) \pa_y^2+ (\beta x+\kappa_1)\pa_x+ (\beta y+\kappa_2)\pa_y$ \\[2mm]
Type IV:  & $L = x \pa_{x}^2+y\pa_{y}^2+(\beta x+\kappa_1)\pa_x+ (\beta y+\kappa_2)\pa_y$ \\[2mm]
Type V:  & $L =  2 x \pa_x\pa_y+ y \pa_y^2+ (\beta x+\kappa_1)\pa_x+ (\beta y+\kappa_2)\pa_y$ \\[2mm]
Type VI:  & $L = x \pa_{x}^2+\pa_{y}^2+ (\beta x+\kappa_1)\pa_x+ (\beta y+\kappa_2)\pa_y$ \\[2mm]
Type VII:  & $L =  \pa_{x}^2+\pa_{y}^2+\beta(x\pa_{x}+x\pa_{y})$ \\[2mm]
Type VIII:  & $L = y \pa_x^2+2 \pa_x\pa_y+  (\beta x+\kappa_1)\pa_x+ (\beta y+\kappa_2)\pa_y$ \\[2mm]
Type IX:  & $L = (x^2-1) \pa_x^2+2 x y \pa_x\pa_y+ (y^2-1) \pa_y^2+ \beta x\pa_x+ \beta y\pa_y$ \\[2mm]
 \hline
\end{tabular}
\end{center}
\end{table}

\br
Requiring {\em only polynomial} coefficients in the form of (\ref{admiss}) {\em guarantees} that the space of polynomials is invariant under the action of $L$.  Requiring polynomial \underline{eigenfunctions} then restricts the degree of these polynomial coefficients.  Requiring that the $N+1$ polynomials (\ref{pmn}) have a common eigenvalue, forces relations between the parameters.

However, {\em specific} families of polynomials can be eigenfunctions of a more general form of operator, as will be seen in Section \ref{sec:cmuv} below.
\er

The matrix of coefficients of the second order terms (if non-degenerate) defines the inverse of a metric, with $L$ being a deformation of the corresponding Laplace-Beltrami operator:
$$
L_b f = \sum_{i,j=1}^2 \frac{1}{\sqrt{g}}\, \frac{\pa}{\pa x^j}\left( \sqrt{g}\, g^{ij}\frac{\pa f}{\pa x^i}\right),
$$
where $g$ is the determinant of the matrix $g_{ij}$ (and with $(x_1,x_2)=(x,y)$).  In \cite{01-1} it was shown that the metrics occurring in the Krall-Sheffer operators were either flat or constant curvature.  The curvature is proportional to the parameter $\alpha$, so the flat cases correspond to $\alpha = 0$ (cases IV, V, VI, VII and VIII).

In \cite{f13-1} we considered the isometry algebras of these 9 metrics.  It is well known that flat and constant curvature metrics possess the maximal group of isometries, which (on a space of dimension $n$) is of dimension $\frac{1}{2}n(n+1)$.  In \cite{f13-1} we construct the 4 symmetry algebras corresponding to the Krall Sheffer metrics, shown in Table \ref{symtab}.
\begin{table}[hbt]
\begin{center}
\caption{The symmetry algebras for the Krall-Sheffer metrics} \label{symtab}\vspace{3mm}
\begin{tabular}{|c|c|}
  \hline
   Types & Symmetry algebra \\ \hline\hline
  I, IX & $so(3)$ \\ \hline
  II, III & $sl(2)$ \\ \hline
  IV, VI, VII & $e(2)$ \\ \hline
  V, VIII & $e(1,1)$ \\
  \hline
\end{tabular}
\end{center}
\end{table}
The constant curvature metrics have isometry algebras $so(3)$ or $sl(2)$. The flat cases have algebras $e(2)$ or $e(1,1)$.  The Laplace-Beltrami operator is just the Casimir of the corresponding algebra.
Isomorphisms between different concrete realisations of a given algebra give {\em non-affine} transformations between corresponding Krall-Sheffer operators.
For example, types II and III are related through the change of coordinates $x_2=-64 x_3,\; y_2=-y_3^2/(4x_3)$.

In \cite{01-1} it was shown that each of the Krall Sheffer operators commute with two other independent differential operators and hence define a superintegrable system.  In \cite{f13-1} we exploited the connection between second (or higher) order commuting operators and Killing vectors.  In the case of flat and constant curvature metrics {\em all} higher order Killing tensors are built from tensor products of Killing vectors.  In our context, this means that the leading order terms of second order, commuting operators are just quadratic expressions in first order differential operators (Killing vectors).  In this paper we illustrate this construction in the context of the Type II Krall-Sheffer operator.

\subsection{The Krall-Sheffer Type II Operator}\label{Sec:KSTypeII}

By looking at the coefficients of the second order part of the operator $L$ in Table \ref{Tab:KSops}, we see that
\begin{subequations}
\be                 \label{KSIImetric}  %
g^{ij} = \left( \begin{array}{cc}
                 x^2 & x y \\
                 x y & y^2-y
                 \end{array}  \right) .
\ee %
This metric has constant curvature and a convenient basis of Killing vectors is   %
\be             \label{sl2-1} %
{\bf H} = 4 x \pa_x,\quad {\bf E} = 2 \sqrt{x y} \pa_y ,\quad
        {\bf F} = 4 \sqrt{x y} \, \pa_x + 2 (y-1) \sqrt{\frac{y}{x}} \, \pa_y ,
\ee %
satisfying the standard commutation relations of $sl(2,\mathbb{C})$:  %
\be                    \label{sl2comm} %
[{\bf H},{\bf E}] = 2 {\bf E}, \quad [{\bf H},{\bf F}] = -2 {\bf F}, \quad
                      [{\bf E},{\bf F}] =  {\bf H} .
\ee %
The Laplace-Beltrami operator for the metric (\ref{KSIImetric}) is proportional to
the quadratic Casimir operator:    %
\be      \label{cas}   %
L_b = \frac{1}{16} ( {\bf H}^2 + 2 {\bf E}{\bf F} + 2 {\bf F}{\bf E}) =
   x^2 \pa_x^2 + 2 x y \pa_x \pa_y + (y^2-y) \pa_y^2 + \frac{3}{2} x \pa_x
                             + \frac{1}{2} (3 y-1) \pa_y  .  %
\ee   %
\end{subequations}
It can be seen that the Krall-Sheffer operator is just a deformation of this, with more general first order coefficients.  Since any quadratic expression in the Killing vectors {\em commutes with $L_b$}, we can similarly deform this to obtain an operator which commutes with $L$.  For example, choosing $I=K^2 + \xi(x,y) \pa_x+\eta(x,y) \pa_y$, where $K$ is some Killing vector, then $[L,I]=0$ gives an over-determined system of partial differential equations for $\xi$ and $\eta$, whose solution gives the form of the operator.  In particular, we have
\begin{subequations}\label{KSiiI12}
\bea
I_1 &=& \frac{1}{16} {\bf H}^2 +  \xi(x,y) \pa_x+\eta(x,y) \pa_y = x^2\pa_x^2+((\beta+\kappa_2)x+\kappa_1(1-y))\pa_x,  \label{KSiiI1}  \\
I_2 &=&  \frac{1}{4} {\bf E}^2 +  \xi(x,y) \pa_x+\eta(x,y) \pa_y =   xy\pa_y^2+(\kappa_1y-\kappa_2x)\pa_y.    \label{KSiiI2}
\eea
\end{subequations}

\subsubsection{Polynomial Eigenfunctions}\label{Sec:KSTypeIIPols}

\begin{figure}[hbt]
\begin{center}
\caption{The triangular lattice of polynomials $P_{m,n}$
with $P_{0,0}=1$.  Horizontal arrows denote the
action of $I_1$ (right) and $I_2$ (left). The operators $r^{(x)}$ and $r^{(y)}$ represent the edge ladder operators.} \label{trilatt}\vspace{6mm}
\unitlength=0.5mm
\begin{picture}(160,80)
\put(80,80){\makebox(0,0){$P_{0,0}$}}
\put(60,60){\makebox(0,0){$P_{1,0}$}}
\put(100,60){\makebox(0,0){$P_{0,1}$}}
\put(40,40){\makebox(0,0){$P_{2,0}$}}
\put(80,40){\makebox(0,0){{$P_{1,1}$}}}
\put(120,40){\makebox(0,0){$P_{0,2}$}}
\put(20,20){\makebox(0,0){$P_{3,0}$}}
\put(60,20){\makebox(0,0){$P_{2,1}$}}
\put(100,20){\makebox(0,0){$P_{1,2}$}}
\put(140,20){\makebox(0,0){$P_{0,3}$}}
\put(0,0){\makebox(0,0){$P_{4,0}$}}
\put(40,0){\makebox(0,0){{$P_{3,1}$}}}
\put(80,0){\makebox(0,0){$P_{2,2}$}}
\put(120,0){\makebox(0,0){{$P_{1,3}$}}}
\put(160,0){\makebox(0,0){$P_{0,4}$}}
\put(75,75){\vector(-1,-1){10}}
\put(85,75){\vector(1,-1){10}}
\put(55,55){\vector(-1,-1){10}}  \put(105,55){\vector(1,-1){10}}
\put(35,35){\vector(-1,-1){10}}  \put(125,35){\vector(1,-1){10}}
\put(15,15){\vector(-1,-1){10}}  \put(145,15){\vector(1,-1){10}}
\put(80,60){\makebox(0,0){$\rightleftarrows$}}
\put(60,40){\makebox(0,0){$\rightleftarrows$}}
\put(100,40){\makebox(0,0){$\rightleftarrows$}}
\put(40,20){\makebox(0,0){$\rightleftarrows$}}
\put(80,20){\makebox(0,0){$\rightleftarrows$}}
\put(120,20){\makebox(0,0){$\rightleftarrows$}}
\put(20,0){\makebox(0,0){$\rightleftarrows$}}
\put(60,0){\makebox(0,0){$\rightleftarrows$}}
\put(100,0){\makebox(0,0){$\rightleftarrows$}}
\put(140,0){\makebox(0,0){$\rightleftarrows$}}
\put(65,75){\makebox(0,0){$r^{(x)}$}}
\put(95,75){\makebox(0,0){$r^{(y)}$}}
\end{picture}
\end{center}
\end{figure}
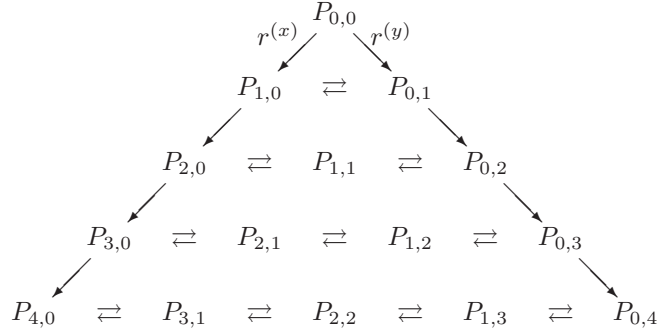

The polynomial eigenfunctions (\ref{pmn}) form a triangular array as depicted in Figure \ref{trilatt}, with $P_{0,0}=1$ at the apex.

The polynomials on the left and right edges of the triangles are functions of only $x$ and $y$ respectively.  These are classical orthogonal polynomials, satisfying {\em 3-point} recurrence relations and possessing first order, ordinary differential ladder operators in each case.  Here we only need to consider the {\em left} edge, with polynomials $P_{m,0}(x)$, which satisfy the eigenvalue problem
\begin{subequations}
\be\label{KSiiLx}
L^{(x)} P_{m,0}=(x^2\pa_x^2+(\beta x+\kappa_1)\pa_x)P_{m,0} = m(m-1+\beta)P_{m,0},
\ee
with 3-point recurrence and ladder operator
\bea
P_{m+1,0} &=& \left(x+\frac{\kappa_1(\beta-2)}{(\beta+2m)(\beta+2m-2)}\right) P_{m,0} \nn\\
&& \hspace{2.5cm} +\left(\frac{m\kappa_1^2(\beta+m-2)}{(\beta+2m-1)(\beta+2m-2)^2(\beta+2m-3)}\right) P_{m-1,0}.  \label{2pmp10}  \\
r^{(x)} &=& \frac{1}{(\beta+2m)(\beta+2m-1)}((\beta+m-1)((\beta+2m)x+\kappa_1)+(\beta+2m)x^2\pa_x).  \label{r+x2}  %
\eea
This ladder operator satisfies the commutation relation
\be\label{KSiiLxr+}
[L^{(x)},r^{(x)}] = \frac{2x(L^{(x)}-\lambda_m)}{\beta+2m-1} +(\lambda_{m+1}-\lambda_m)r^{(x)}, \quad\mbox{with}\quad \lambda_m = m(m-1+\beta).
\ee
\end{subequations}
We can use these to build polynomial eigenfunctions $P_{m,0}$, the first few of which are:
\bea
&&  P_{0,0}=1,\quad P_{1,0}=x+\frac{\kappa_1}{\beta},\quad P_{2,0}=x^2+\frac{2\kappa_1 x}{\beta+2}+\frac{\kappa_1^2}{(\beta+2)(\beta+1)}, \nn\\[-2mm]
   &&   \label{KSiiPm0}    \\[-2mm]
&&  P_{3,0} = x^3 + \frac{3\kappa_1 x^2}{\beta+4} +\frac{3\kappa_1^2 x}{(\beta+4)(\beta+3)} +\frac{\kappa_1^3}{(\beta+4)(\beta+3)(\beta+2)} .  \nn
\eea
By considering the action of $I_k$ on $x^my^n$ it is possible to deduce that
\bea  %
I_1P_{m,n}-m(\beta+\kappa_2+m-1)P_{m,n} &=& -m\kappa_1P_{m-1,n+1} ,\nn\\
I_2P_{m,n}-n\kappa_1P_{m,n} &=& -n(\kappa_2-n+1)P_{m+1,n-1} ,\nn
\eea  %
We see that, starting with $P_{N,0}$, $I_1$ moves us to the right by constructing $P_{m,n}$, with $m+n=N$, until we reach $P_{0,N}$, which satisfies $I_1P_{0,N}=0$.  $I_2$ similarly moves us from right to left.
For example, with $N=3$, we construct
\bea
 P_{2,1} &=& x^2y + \frac{1}{\beta+4}\left(\kappa_2 x^2+2 \kappa_1 x y\right) +\frac{\kappa_1}{(\beta+4)(\beta+3)}\left(2\kappa_2 x+\kappa_1 y\right) +\frac{\kappa_1^2\kappa_2}{(\beta+4)(\beta+3)(\beta+2)} ,  \nn\\
 P_{1,2} &=& x y^2 + \frac{1}{\beta+4}\left(2(\kappa_2-1) x y+ \kappa_1 y^2\right) +\frac{\kappa_2-1}{(\beta+4)(\beta+3)}\left(\kappa_2 x+2\kappa_1 y\right)
                                           +\frac{\kappa_1\kappa_2(\kappa_2-1)}{(\beta+4)(\beta+3)(\beta+2)} ,  \nn\\
 P_{0,3} &=& y^3 + \frac{3(\kappa_2-2)y^2}{\beta+4} + \frac{3(\kappa_2-2)(\kappa_2-1)y}{(\beta+4)(\beta+3)}+\frac{(\kappa_2-2)(\kappa_2-1)\kappa_2}{(\beta+4)(\beta+3)(\beta+2)} .  \nn
\eea

\br
It is possible to build ``internal'' ladder operators, which take us {\em parallel} to the edges, but these are very complicated second order, {\em partial} differential operators (see \cite{f13-1}).  Similarly, there are 3 \underline{level} recurrence relations, which, in this case, connect 6 polynomials in an ``inverted triangle''.
\er

\section{Krall-Sheffer Operators in 3 Dimensions}\label{Sec:KS3D}

We now consider operators in 3 dimensions, of the form
\begin{subequations}
\be\label{gen3D}
L = A_{11} \pa_x^2 + A_{22} \pa_y^2 +A_{33} \pa_z^2  +2A_{12} \pa_x\pa_y +2A_{13} \pa_x\pa_z+2A_{23}\pa_y\pa_z + A_1 \pa_x+A_2 \pa_y+A_3 \pa_z,
\ee
possessing {\em polynomial eigenfunctions}.  Generally, the coefficients should be of the same form as (\ref{admiss}), but we specifically require that on each of the 2 dimensional coordinate surfaces, the operator should reduce to one of standard Krall-Sheffer ones of Table \ref{Tab:KSops}.  There are just 7 consistent combinations (see \cite{scott}), but here we only consider one of these
\bea
L &=& x^2 \pa_x^2 + (y^2-y) \pa_y^2 + (z^2-z) \pa_z^2 +2 \left(x y \pa_x\pa_y  + x z \pa_x\pa_z  + y z \pa_y\pa_z\right) \nn\\
   &&  \hspace{4cm}   + (\beta x+\kappa_1)\pa_x+ (\beta y+\kappa_2)\pa_y + (\beta z+\kappa_3)\pa_z,   \label{3DKS122}
\eea
\end{subequations}
which reduces to Krall-Sheffer Type II on both the $(x,y)$ and $(x,z)$ planes, and to Type I on the $(y,z)$ plane.

In this Section, we present the isometry algebra of the corresponding metric and then 4 independent commuting operators.  We then consider the construction of polynomial eigenfunctions, by first taking the 2D polynomials in $(x,y)$, found in Section \ref{Sec:KSTypeIIPols}, and then applying the commuting operators in 3D.

\subsection{The Isometry Algebra and Commuting Operators}\label{Sec:KS3D-Isom}

The metric corresponding to the coefficients of the second order part of (\ref{3DKS122}) has constant curvature, with a 6 dimensional isometry group, with generators:
\begin{subequations}
\bea
&&  e_1 = 2 \sqrt{xz}\, \pa_z,\;\;\; h_1= 4 x\,\pa_x,\;\;\; f_1 =  2(z+y-1)\sqrt{\frac{z}{x}}\,\pa_z+4\sqrt{x z}\,\pa_x, \nn   \\[-3mm]
  &&   \label{KS3D-g1g2}   \\[-3mm]
&&  e_2 = 2\sqrt{x y}\,\pa_y,\;\;\; h_2 = 4\sqrt{y z} (\pa_y-\pa_z),\;\;\; f_2 = 8\sqrt{x y}\,\pa_x + 4(z+y-1)\sqrt{\frac{y}{x}}\,\pa_y,  \nn
\eea
satisfying the commutation relations of Table \ref{Tab:g1g2}.
\begin{table}[h]
\begin{center}
\caption{The commutation relations $[X,Y]$ of the isometry algebra (\ref{KS3D-g1g2})}\label{Tab:g1g2}\vspace{3mm}
\begin{tabular}{|c|ccc|ccc|}
\hline
     &$e_1$    &$h_1$  &$f_1$     &$e_2$   &$h_2$   &$f_2$\\[.10cm]\hline
$e_1$ &$0$    &$-2 e_1$  &$h_1$    &$0$    &$2 e_2$     &$-2 h_2$\\
$h_1$ &       &$0$     &$-2 f_1$   &$2 e_2$    &$0$  &$-2 f_2$\\
$f_1$ &&               &$0$       &$h_2$       &$f_2$  &$0$\\\hline
$e_2$ &&&     &$0$        &$-2 e_1$ &$2 h_1$ \\
$h_2$ &&&&    &$0$        &$4 f_1$       \\
$f_2$ &&&&&   &$0$\\
\hline
\end{tabular}
\end{center}
\end{table}

This algebra has two quadratic Casimirs:
\bea
L_b &=& \frac{1}{16} \left(2 e_1f_1+2 f_1 e_1+h_1^2+e_2 f_2+f_2 e_2- h_2^2\right)  \nn\\
  &=& x^2 \pa_x^2 + (y^2-y) \pa_y^2 + (z^2-z) \pa_z^2 +2 \left(x y \pa_x\pa_y  + x z \pa_x\pa_z  + y z \pa_y\pa_z\right) \nn\\
   &&  \hspace{4cm}   + 2 x\pa_x+ \left(2 y-\frac{1}{2}\right)\pa_y +  \left(2 z-\frac{1}{2}\right)\pa_z,   \label{3DKSLb}
\eea
and
\be
{\cal C}_2 = 2 h_1 h_2 +e_1 f_2+f_2 e_1-2 e_2 f_1-2 f_1 e_2 \equiv 0.
\ee
\end{subequations}
The first of these is just the Laplace-Beltrami operator of our metric, and the second is a quadratic constraint on the algebra (in this realization).  The operator (\ref{3DKS122}) is just a deformation of $L_b$.

In a constant curvature space, {\em all} higher order Killing tensors are built from tensor products of Killing vectors.  This means that higher order operators, commuting with $L_b$, are just built from products of the Killing vectors (first order operators).  For example, the operator $e_1^2 = 2 x z \pa_x^2+2 z \pa_x$ commutes with $L_b$.  We can add some first order ``correction terms'' to obtain an operator which commutes with $L$ (of (\ref{3DKS122})).  We build 4 such operators, commuting with $L$, by deforming (some multiples of) $e_1^2,\, e_2^2,\, h_1^2$ and $h_2^2$, respectively:
\begin{subequations}
\bea
 I_1 &=& xz\pa_z^2+(\kappa_1 z-\kappa_3 x) \pa_z,   \label{3DKS-I1}    \\
 I_2 &=& x y\pa_y^2+(\kappa_1 y-\kappa_2 x) \pa_y,   \label{3DKS-I2}    \\
  I_3 &=& x^2\pa_x^2+((1-z-y)\kappa_1+(\beta+\kappa_2+\kappa_3) x) \pa_x,   \label{3DKS-I3}  \\
  I_4 &=& y z (\pa_y-\pa_z)^2+(\kappa_3 y-\kappa_2 z) (\pa_y-\pa_z). \label{3DKS-I4}
\eea
\end{subequations}
Since $[I_1,I_2]=[I_3,I_4]=0$, we can use either of the triples $\{L,I_1,I_2\}$ or $\{L,I_3,I_4\}$ to show that the system is {\em completely integrable}, whilst the existence of all 4 operators, commuting with $L$, shows that the system is {\em maximally superintegrable}.

\subsection{The Polynomial Eigenfunctions}\label{Sec:KS3D-Pols}

We now consider a basis of \underline{monic} {\em polynomial eigenfunctions} of the form
\be \label{Plmn}
P_{\ell,m,n} (x,y,z) = x^{\ell}y^m z^n+\mbox{lower order terms},\quad 0\leq \ell,m,n.
\ee
These form a {\em tetrahedral array}, with $P_{0,0,0}=1$ at its apex.  On each triangular face we have polynomials in {\em two} variables, which are eigenfunctions of a 2D Krall-Sheffer operator (of either Type I or Type II).
In particular, on the $x-y$ face we build polynomials $P_{\ell,m,0}(x,y)$, which are just the eigenfunctions we built in Section \ref{Sec:KSTypeIIPols}.

Recall that we used a {\em ladder operator} to build a sequence of 1 dimensional polynomials $P_{N,0,0}(x)$, with eigenvalue $\lambda_{N} = N(N-1+\beta)$.  The operator $I_1$ (of (\ref{KSiiI1})) then allowed us to build $N+1$ eigenfunctions $\{P_{\ell,m,0}(x,y)\}_{\ell+m=N}$, with the same eigenvalue.

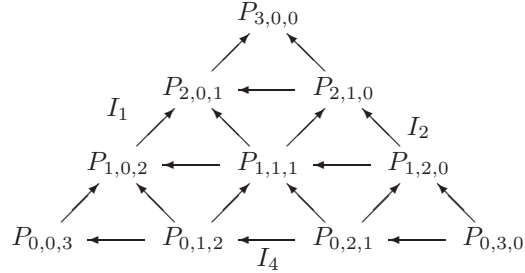
\begin{figure}[hbt]
\begin{center}
\caption{The horizontal triangular lattice of polynomials $P_{\ell,m,n}$ with $\ell+m+n=3$.  Horizontal arrows to the left depict the action of $I_4$. Arrows parallel to the left (right) edges correspond to $I_1$ ($I_2$)} \label{horiz-triangle}\vspace{6mm}
\unitlength=0.5mm
\begin{picture}(140,60)
\put(60,60){\makebox(0,0){$P_{3,0,0}$}}
\put(40,40){\makebox(0,0){$P_{2,0,1}$}}
\put(80,40){\makebox(0,0){$P_{2,1,0}$}}
\put(20,20){\makebox(0,0){$P_{1,0,2}$}}
\put(60,20){\makebox(0,0){{$P_{1,1,1}$}}}
\put(100,20){\makebox(0,0){$P_{1,2,0}$}}
\put(0,0){\makebox(0,0){$P_{0,0,3}$}}
\put(40,0){\makebox(0,0){$P_{0,1,2}$}}
\put(80,0){\makebox(0,0){$P_{0,2,1}$}}
\put(120,0){\makebox(0,0){$P_{0,3,0}$}}
\put(45,45){\vector(1,1){10}}
\put(75,45){\vector(-1,1){10}}
\put(25,25){\vector(1,1){10}}  \put(65,25){\vector(1,1){10}}
\put(5,5){\vector(1,1){10}}  \put(95,25){\vector(-1,1){10}}  \put(55,25){\vector(-1,1){10}}
  \put(115,5){\vector(-1,1){10}}  \put(75,5){\vector(-1,1){10}}  \put(35,5){\vector(-1,1){10}}
  \put(45,5){\vector(1,1){10}} \put(85,5){\vector(1,1){10}}
\put(67,40){\vector(-1,0){15}}
\put(87,20){\vector(-1,0){15}}
\put(47,20){\vector(-1,0){15}}
\put(107,0){\vector(-1,0){15}}
\put(67,0){\vector(-1,0){15}}
\put(27,0){\vector(-1,0){15}}

\put(20,35){\makebox(0,0){$I_1$}}
\put(100,30){\makebox(0,0){$I_2$}}
\put(60,-5){\makebox(0,0){$I_4$}}
\end{picture}
\end{center}
\end{figure}

We now show how to build a further $\frac{1}{2} N (N+1)$ polynomial eigenfunctions of the form (\ref{Plmn}), giving us a total of $\frac{1}{2} (N+1)(N+2)$, having the same eigenvalue $\lambda_{N}$.  They form a horizontal, triangular slice of the full tetrahedral array of eigenfunctions.  This is depicted in Figure \ref{horiz-triangle}, for the case $N=3$.

By looking at the action of $I_i$ on the leading term of $P_{\ell,m,n}$, we can determine the action on polynomials (\ref{Plmn}):
\begin{subequations}
\bea
\hspace{-3mm} I_1 P_{\ell,m,n} &=&  \kappa_1 n P_{\ell,m,n} + n (n-1-\kappa_3) P_{\ell+1,m,n-1},  \label{KS3D-I1P}  \\
\hspace{-3mm} I_2 P_{\ell,m,n} &=&  \kappa_1 m P_{\ell,m,n} + m (m-1-\kappa_2) P_{\ell+1,m-1,n},  \label{KS3D-I2P}  \\
\hspace{-3mm} I_3 P_{\ell,m,n} &=&  \ell (\ell-1+\beta+\kappa_2+\kappa_3) P_{\ell,m,n} -\ell \kappa_1 (P_{\ell-1,m+1,n}+P_{\ell-1,m,n+1}),  \label{KS3D-I3P}  \\
\hspace{-3mm} I_4 P_{\ell,m,n} &=&  (\kappa_3 m+\kappa_2 n-2 m n) P_{\ell,m,n} + n (n-1-\kappa_3) P_{\ell,m+1,n-1}+ m (m-1-\kappa_2) P_{\ell,m-1,n+1}.  \label{KS3D-I4P}
\eea
\end{subequations}
First note that the action of $I_4$ is to keep $\ell$ fixed, whilst {\em increasing} $n$ and decreasing $m$.  When $n=0$, the coefficient of $P_{\ell,m+1,n-1}$ vanishes, so this potentially contradictory term does not arise.  For $m \geq 1$, (\ref{KS3D-I4P}) can be solved for $P_{\ell,m-1,n+1}$.  When $m=0$, the coefficient of $P_{\ell,m-1,n+1}$ vanishes, so again this potentially contradictory term does not arise.  This procedure is indicated by the horizontal arrow going to the {\em left} in Figure \ref{horiz-triangle}.  Since we already know the polynomials $P_{\ell,m,0}$ on the right edge, we use $I_4$ to build the remaining polynomials in Figure \ref{horiz-triangle}.  The operators $I_1$ and $I_2$ also act in a simple way on these polynomials, with $I_1$ keeping $m$ fixed and moving upwards and parallel to the {\em left edge}.  Similarly, $I_2$ keeps $n$ fixed and moves us upwards, but parallel to the {\em right edge}.

Before writing down the explicit polynomials depicted in Figure \ref{horiz-triangle}, note the following {\em discrete symmetry}:
\begin{subequations}
\be\label{yzSymm}
\iota_{yz}: (x,y,z,\kappa_1,\kappa_2,\kappa_3) \mapsto (x,z,y,\kappa_1,\kappa_3,\kappa_2) \quad\Rightarrow\quad L\mapsto L \;\;\;\mbox{and}\;\;\; P_{\ell,m,n}  \mapsto  P_{\ell,n,m}.
\ee
For our case of $N=3$, this means that the left edge polynomials, $P_{2,0,1},\,P_{1,0,2},\, P_{0,0,3}$, are derived from their counterparts on the right edge.  Furthermore, $P_{0,1,2}$ is similarly related to $P_{0,2,1}$ and $P_{1,1,1}$ is {\em invariant}.  We, therefore, just need to derive two polynomials:
\bea
P_{1,1,1} &=& x y z + \frac{\kappa_3 x y+ \kappa_2 x z+\kappa_1 y z}{\beta+4} + \frac{\kappa_2\kappa_3 x
                 + \kappa_1\kappa_3 y +\kappa_1\kappa_2 z}{(\beta+4)(\beta+3)} + \frac{\kappa_1\kappa_2\kappa_3}{(\beta+4)(\beta+3)(\beta+2)} , \label{p111}  \\
P_{0,2,1} &=& y^2 z + \frac{\kappa_3 y^2 +2 (\kappa_2-1) y z}{\beta+4} + \frac{(\kappa_2-1) (2\kappa_3 y + \kappa_2 z)}{(\beta+4)(\beta+3)} + \frac{\kappa_2(\kappa_2-1)\kappa_3}{(\beta+4)(\beta+3)(\beta+2)} . \label{p021}
\eea
\end{subequations}

\br
This is the first value of $N$ for which there is an ``internal'' polynomial, $P_{1,1,1}$, depending upon {\em all three} variables.  For $N\geq 3$, we have $\frac{1}{2}(N-1)(N-2)$ such internal polynomials.
\er

\br
The involution $y\leftrightarrow z$ acts as an automorphism on the algebra (\ref{KS3D-g1g2}):
$$
(e_1,h_1,f_1,e_2,h_2,f_2) \mapsto \left(e_2,h_1,\frac{1}{2}f_2,e_1,-h_2,2f_1\right),
$$
so naturally preserves the Casimir $L_b$.  Extending this to the parameters $\kappa_i$, to give $\iota_{yz}$, of (\ref{yzSymm}), then naturally preserves $L$, and acts on the commuting operators $I_i$, with  $\iota_{yz}: (I_1,I_2,I_3,I_4)\mapsto (I_2,I_1,I_3,I_4)$.
\er

\section{The Rational Calogero-Moser System with Harmonic Potential (2 Dimensional Case)}\label{Sec:2D-CM}

The rational Calogero-Moser system (both classical and quantum) \cite{71-3} has been fundamental in the history of integrable systems, and is well known to be superintegrable \cite{83-11,96-5}.  This continues to be the case when we add an isotropic harmonic oscillator potential \cite{97-9,13-3}.  There is a large choice of commuting operators, but here we consider the Calogero-Moser term to be a deformation of the isotropic oscillator:
\begin{subequations}
\bea
&&  L = \pa_1^2+\pa_2^2-\omega^2 \left(q_1^2+ q_2^2\right) - \frac{2 c^2}{(q_1-q_2)^2},  \label{Lcm}  \\
&&  K_1 =  \left(q_2\pa_1-q_1\pa_2\right)^2 -\frac{4 c^2 q_1q_2}{(q_1-q_2)^2}  ,  \quad K_2 =  \pa_1\pa_2 - \omega^2 q_1q_2+\frac{c^2}{(q_1-q_2)^2} ,  \label{K12cm}
\eea
where $\pa_i=\pa_{q_i}$.  These satisfy $[L,K_i]=0$, but $K_3=[K_1,K_2]$ is a third order operator, satisfying
\bea
  \left[ K_1,K_3 \right] &=& - 8 \left(K_1K_2+K_2K_1+2 (1-2 c^2) K_2-c^2 L\right),  \label{K1K3cm}  \\
  \left[ K_2,K_3 \right] &=& 2 \left(4 K_2^2-L^2-8 \omega^2 K_1 +4 (1+2 c^2) \omega^2\right),  \label{K2K3cm}  \\
  K_3^2 &=& 4 \left(K_1L^2-\frac{4}{3}\left(K_2^2K_1+K_2K_1K_2+K_1K_2^2\right)  +\frac{4}{3} (6c^2-11)K_2^2+4 \omega^2 K_1^2  \right.  \nn\\
&&       \left. \quad  -4 c^2 K_2 L +\frac{2}{3} L^2-\frac{4}{3}\omega^2(6c^2+11) K_1  +\frac{4}{3} \omega^2(2-20c^2-3 c^4)\right).  \label{K32cm}
\eea
\end{subequations}
The element $K_3$ is not needed in what follows, but these relations show that there is no need to add further elements to close the algebra.  The algebraic constraint (\ref{K32cm}) reflects the fact that we can only have 3 independent commuting operators in 2D.

We next introduce the gauge transformation $L\mapsto \hat L = g L g^{-1}$, with $g = (q_1-q_2)^p e^{\frac{1}{2} \omega (q_1^2+q_2^2)}$, to obtain
\begin{subequations}
\bea
  \hat L &=& \pa_1^2+\pa_2^2-2\omega \left(q_1\pa_1+ q_2\pa_2\right) - \frac{2 p}{(q_1-q_2)}\left(\pa_1-\pa_2\right)+2 (p-1)\omega ,  \label{Lhatcm}  \\
         \hat K_1 &=&  \left(q_2\pa_1-q_1\pa_2\right)^2 -\frac{2 p (q_1+q_2)}{(q_1-q_2)} \left(q_2\pa_1-q_1\pa_2\right) + p (p+2) ,  \label{K1hatcm} \\
  \hat K_2 &=&  \pa_1\pa_2 - \omega \left(q_2\pa_1+ q_1\pa_2\right)+\frac{p}{(q_1-q_2)}\left(\pa_1-\pa_2\right)-p\omega,  \label{K2hatcm}
\eea
\end{subequations}
when $c^2= p (p+1)$.

Finally, the orthogonal transformation, $u=\frac{1}{\sqrt{2}} (q_1+q_2),\, v=\frac{1}{\sqrt{2}} (q_1-q_2)$, gives
\begin{subequations}
\bea
  \hat L &=& \pa_u^2+\pa_v^2-2\omega \left(u\pa_u+ v\pa_v\right) - \frac{2 p}{v}\pa_v+2 (p-1)\omega ,  \label{Luvcm}  \\
          \hat K_1 &=&  \left(v\pa_u-u\pa_v\right)^2 -\frac{2 p u}{v} \left(v\pa_u-u\pa_v\right) + p (p+2) ,  \label{K1uvcm} \\
  \hat K_2 &=&  \frac{1}{2} \left(\pa_u^2-\pa_v^2\right) - \omega \left(u\pa_u- v\pa_v\right)+\frac{p}{v}\pa_v -p\omega.  \label{K2uvcm}
\eea
\end{subequations}

In these coordinates (the non-constant part of) $\hat L$ separates into $\hat L = L^{(u)}+L^{(v)}$, satisfying $[L^{(u)},L^{(v)}]=0$.  Then, $\hat K_1 = \frac{1}{2} \left(L^{(u)}-L^{(v)}\right)$.

\subsection{Building Eigenfunctions}\label{sec:cmuv}

We again construct a triangular array of eigenfunctions.  On the left edge, eigenfunctions depend upon $u$ only and consist of Hermite polynomials $P_{m,0},\; m\geq 0$, with eigenvalue $\lambda_m=-2 m \omega$.  The commuting operators allow us to build other polynomial eigenfunctions with the same eigenvalue.  On the right edge, we also have polynomial eigenfunction, just in the variable $v$, but only of even degree.

\subsubsection{The Left Edge}\label{sec:Hermite}

We consider the 1-dimensional eigenvalue problem, with corresponding ladder operator:
\begin{subequations}\label{LAucm}
\be \label{Lucm}
L^{(u)}P_{m,0} = (\pa_u^2-2 \omega \pa_u)P_{m,0} = \lambda_m P_{m,0},\quad A_u^+ P_{m,0} = \left(u -\frac{1}{2\omega}\, \pa_u\right)P_{m,0},
\ee
satisfying
\be \label{ladcm}
[L^{(u)},A_u^+] = -2\omega A_u^+ \quad \Rightarrow\quad  \lambda_m=-2 m \omega,
\ee
when $P_{0,0}=1$.  We therefore have Hermite polynomials on the left edge:
\be \label{HPolscm}
P_{1,0}=u,\quad P_{2,0}=u^2-\frac{1}{2\omega},\quad P_{3,0} = u \left(u^2-\frac{3}{2\omega}\right),\quad P_{4,0}=u^4-\frac{3}{\omega}\, u^2+\frac{3}{4\omega^2}, \dots
\ee
\end{subequations}
These are the polynomials briefly discussed in Section \ref{Sec:ResHarmonicHermite}.

\subsubsection{Using the Commuting Operators}\label{sec:pols-cm}

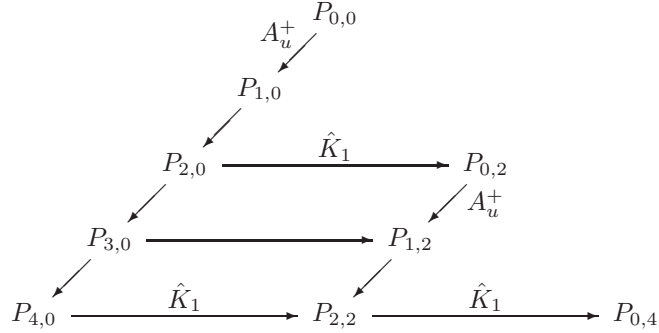
\begin{figure}[hbt]
\begin{center}
\caption{The triangular lattice, {\em partially} filled by polynomials $P_{m,n}$, with $P_{0,0}=1$.  Horizontal arrows denote the
action of $\hat K_1$ and parallel to the left edge we have $A_u^+$.} \label{uvlatt}\vspace{6mm}
\unitlength=0.5mm
\begin{picture}(160,80)
\put(80,80){\makebox(0,0){$P_{0,0}$}}
\put(60,60){\makebox(0,0){$P_{1,0}$}}
\put(40,40){\makebox(0,0){$P_{2,0}$}}
\put(120,40){\makebox(0,0){$P_{0,2}$}}
\put(20,20){\makebox(0,0){$P_{3,0}$}}
\put(100,20){\makebox(0,0){$P_{1,2}$}}
\put(0,0){\makebox(0,0){$P_{4,0}$}}
\put(80,0){\makebox(0,0){$P_{2,2}$}}
\put(160,0){\makebox(0,0){$P_{0,4}$}}
\put(75,75){\vector(-1,-1){10}}
\put(55,55){\vector(-1,-1){10}}
\put(35,35){\vector(-1,-1){10}}
\put(15,15){\vector(-1,-1){10}}
\put(115,35){\vector(-1,-1){10}}
\put(95,15){\vector(-1,-1){10}}
\put(50,40){\vector(1,0){60}}
\put(30,20){\vector(1,0){60}}
\put(10,0){\vector(1,0){60}}
\put(90,0){\vector(1,0){60}}
\put(65,75){\makebox(0,0){$A_u^+$}}
\put(120,30){\makebox(0,0){$A_u^+$}}
\put(80,45){\makebox(0,0){$\hat K_1$}}
\put(40,5){\makebox(0,0){$\hat K_1$}}
\put(120,5){\makebox(0,0){$\hat K_1$}}
\end{picture}
\end{center}
\end{figure}

We use the operator $\hat K_1$ of (\ref{K1uvcm}) (but drop the constant term $p(p+2)$) to build eigenfunctions to the right of $P_{m,0}$.  Considering the action $\hat K_1(u^mv^n)$ we find
\begin{subequations}
\be \label{rstep-cm}
\hat K_1P_{m,n} = n(n-1-2 p) P_{m+2,n-2} -\left((1-2 p)m +(1+2 m)n\right) P_{m,n} + m (m-1) P_{m-2,n+2}.
\ee
In particular
$$
\hat K_1P_{m,0} =  -(1-2 p)m P_{m,0} + m (m-1) P_{m-2,2}\quad\mbox{and}\quad \hat K_1P_{0,n} = n(n-1-2 p) P_{2,n-2} -2 n P_{m,n},
$$
so we never encounter the terms to left and right of our array.    Furthermore
$$
\hat K_1(u^m)=m(2 p-1)u^m+m(m-1)u^{m-2}v^2,
$$
so only \underline{positive and even} powers of $v$ occur (despite the $v^{-1}$ term in the formula).

Starting from either $m=2N$ or $m=2 N+1$, on the left edge, we generate $N$ polynomial eigenfunctions $\{P_{2(N-k),2 k}\}_{k=0}^{N}$ or $\{P_{2(N-k)+1,2 k}\}_{k=0}^{N}$.  Furthermore, since the eigenvalue problem is {\em separable} in $(u,v)$ coordinates, the eigenfunctions are just {\em products}:
\be \label{Pmn-cm-sols}
P_{2(N-k),2 k}= P_{2(N-k),0}\, P_{0,2 k} \quad\mbox{and}\quad P_{2(N-k)+1,2 k} = P_{2(N-k)+1,0}\, P_{0,2 k}.
\ee
We therefore only need to enumerate the polynomials on the left and right edges and then to apply the above formulae.  The first 4 polynomials on the {\em left} edge are listed in (\ref{HPolscm}), with the first 3 polynomials on the right edge being:
\bea
&&  P_{0,2}= v^2+\frac{2 p-1}{2\omega},\quad P_{0,4}=v^4+\frac{(2 p-3)}{\omega}\, v^2+\frac{(2 p-3)(2 p-1)}{4\omega^2},\nn\\[-2mm]
&&            \label{P0n-cm-sols}   \\[-2mm]
        &&           P_{0,6}= v^6+\frac{3(2 p-5)}{2\omega}\, v^4+\frac{3(2 p-5)(2 p-3)}{4\omega^2} \, v^2 + \frac{(2 p-5)(2 p-3)(2 p-1)}{8\omega^3}.  \nn
\eea
\end{subequations}
This array of polynomials {\em partially} fills a triangular lattice, as shown in Figure \ref{uvlatt}.  When $p=0$, these are just the {\em even} Hermite polynomials of Section \ref{Sec:ResHarmonicHermite}.

Our route to $P_{m,n}$ was to use the ladder operator $A_u^+$ to construct $P_{m,0}$ and then to build the other eigenfunctions, with a given eigenvalue, by operating with $\hat K_1$.  However, as indicated in Figure \ref{uvlatt}, we can use $A_u^+$ to build $P_{m,2n}$, with $m\geq 1$, from $P_{0,2n}$.

\subsubsection{Returning to the Original Coordinates}

Taking any $P_{m,n}(u,v)$, we obtain an eigenfunction of the original operator (\ref{Lcm}) by the transformation
\be \label{phimn}
\phi_{m,n}(q_1,q_2) = (q_1-q_2)^{-p}\, P_{m,n}\left(\frac{q_1+q_2}{\sqrt{2}},\frac{q_1-q_2}{\sqrt{2}}\right)\,  e^{-\frac{1}{2} \omega (q_1^2+q_2^2)}.
\ee
We add the constant $2(p-1)\omega$, from (\ref{Luvcm}), to the eigenvalue $-2m\omega$, from (\ref{ladcm}), to obtain $\lambda_{m,n}=2(p-1-m-n)\omega$.

{\em In this way we have \underline{directly} used the superintegrability to build a lattice of eigenfunctions of the operator (\ref{Lcm}) for the Calogero-Moser system in 2 dimensions.}

\section{The Rational Calogero-Moser System with Harmonic Potential (3 Dimensional Case)}\label{Sec:3D-CM}

We now consider
\begin{subequations}
\be \label{Lcm3D}
L = \pa_1^2+\pa_2^2+\pa_3^2-\omega^2 \left(q_1^2+ q_2^2+q_3^2\right) - 2 c^2 \sum_{i< j}  \frac{1}{(q_i-q_j)^2},
\ee
where $\pa_i=\pa_{q_i}$, and which we again consider as a deformation of the isotropic oscillator.  The isotropic oscillator is rotationally invariant, with angular momentum operators (labelled modulo 3)
\be \label{Omega}
\Omega_i = q_{i-1} \pa_{i+1} - q_{i+1} \pa_{i-1},\;\;i=1,\dots ,3, \quad\mbox{with Casimir}\;\; \Omega^2=\Omega_1^2+\Omega_2^2+\Omega_3^2.
\ee
Whilst (\ref{Lcm3D}) is not rotationally invariant, we can deform the Casimir to obtain a commuting operator:
\be \label{Omega2}
K_1 = \Omega^2 -2 c^2 \sum_{i< j}\, \frac{q_{6-i-j}^2+2 q_i q_j}{(q_i-q_j)^2},
\ee
\end{subequations}
which will play the role of $K_1$ in (\ref{K12cm}).  There are more commuting operators (the system is superintegrable), but we only use this one here.

The gauge transformation in 3 dimensions is $L\mapsto \hat L = g L g^{-1}$, with
$$
g = \left((q_1-q_2)(q_2-q_3)(q_3-q_1)\right)^p e^{\frac{1}{2} \omega (q_1^2+q_2^2+q_3^2)},
$$
giving (labelling modulo 3)
\begin{subequations}
\bea
  \hat L &=& \sum_{i=1}^3 \left(\pa_i^2-2\omega q_i\pa_i \right) - 2 p \sum_{i< j}\, \frac{\left(\pa_i-\pa_j\right)}{(q_i-q_j)}+ 3 (2p-1)\omega ,  \label{Lhatcm3D}  \\
         \hat K_1 &=&  \Omega^2 - 2 p \sum_{i=1}^3 \left(q_{i+2}+q_{i+1}-q_i + \frac{2 q_{i+1}^2+q_{i+2}^2}{q_i-q_{i+1}} + \frac{q_{i+1}^2+2q_{i+2}^2}{q_i-q_{i+2}}\right)\, \pa_i - 3 p (p-3) ,  \label{K1hatcm3D}
\eea
\end{subequations}
when $c^2= p (p+1)$.

The orthogonal transformation
\begin{subequations}
\be\label{uvw}
u=\frac{1}{\sqrt{3}} (q_1+q_2+q_3),\quad v=\frac{1}{\sqrt{6}} (q_1-2 q_2+q_3), \quad w = \frac{1}{\sqrt{2}} (q_1-q_3)
\ee
then gives (removing the constant terms)
\bea
  \hat L &=& \pa_u^2+\pa_v^2+\pa_w^2-2\omega \left(u\pa_u+ v\pa_v+w\pa_w\right) - \frac{6 p}{3 v^2-w^2}\, \left(2 v \pa_v+(v^2-w^2) \pa_w\right) ,  \label{Luvcm3D}  \\
  \hat K_1 &=&  \Omega^2 +6 p (u\pa_u+3v\pa_v+w\pa_w)  \nn\\
  &&   \hspace{2cm} - \frac{6 p}{w (3v^2-w^2)}\, \left( 2 v w (u^2+4 v^2)\pa_v+(v^2-w^2)(u^2+v^2+w^2)\pa_w\right). \label{K1uvcm3D}
\eea
\end{subequations}

In these coordinates $\hat L$ separates into $\hat L = L^{(u)}+L^{(v,w)}$, satisfying $[L^{(u)},L^{(v,w)}]=0$.

\subsection{Building Eigenfunctions}\label{sec:cmuv3D}

We again construct a {\em triangular array} of eigenfunctions ({\em not} a pyramid).  On the left edge, $\hat L$ reduces to $L^{(u)}$, whose eigenfunctions are exactly the Hermite polynomials of Section \ref{sec:Hermite}.

Using $\hat K_1$ to construct other eigenfunctions we see that,
$$
\hat K_1 \left(u^m\right) = 2 m (3 p-1) u^m +m(m-1) u^{m-2} \left(v^2+w^2\right) \quad\Rightarrow\quad \hat K_1 \left(P_{m,0}(u)\right) = {\cal P}\left(u,v^2+w^2\right),
$$
for some polynomial $\cal P$ of two variables.  To determine the next step in the calculation, we transform to polar coordinates in the $(v,w)$ plane: $v=r \cos\theta,\, w = r \sin \theta$
\begin{subequations}
\bea
\hat L  &=& \pa_u^2-2 \omega u \pa_u +\pa_r^2 -\left(2 \omega r +\frac{6 p-1}{r}\right)\, \pa_r +\frac{1}{r^2} \left(\pa_\theta^2-6 p \cot 3 \theta \, \pa_\theta\right) ,  \label{Lurt}  \\
\hat K_1  &=& \left(r \pa_u-u \pa_r\right)^2+ (6p-1)\, \frac{u}{r}\, \left(r \pa_u-u \pa_r\right) +\left(1+\frac{u^2}{r^2}\right) \left(\pa_\theta^2-6 p \cot 3 \theta \, \pa_\theta\right). \label{K1urt}
\eea
\end{subequations}
As written, it is clear that $\hat L$ {\em separates} in these coordinates.  Furthermore, the space of functions of the two variables $(u,r)$ is {\em invariant} under the action of {\em both} operators.  Thus, starting with a function of $u$ only (on the left edge) we build functions of only the 2 variables $(u,r)$.  Furthermore, when acting on $f(u,r)$, these operators reduce to {\em exactly} those of (\ref{Luvcm}) and (\ref{K1uvcm}), but with $(v,p)$ replaced by $\left(r,\frac{1}{2} (6p-1)\right)$.

Therefore, our previous 2 dimensional polynomials give us polynomial solutions of $(u,r)$, which only contain {\em even} powers of $r$, so correspond to {\em polynomials} in $(u,v,w)$.  Specifically, if we label our polynomials of Section \ref{sec:pols-cm} as $P_{m,n}(p,u,v)$ then
$$
{\cal P}_{m,n}(p,u,r)=P_{m,n}\left(\frac{1}{2} (6p-1),u,r\right) \quad\mbox{satisfies}\quad  \hat L {\cal P}_{m,n}(p,u,r) = -2m \omega {\cal P}_{m,n}(p,u,r).
$$

\subsection{Returning to the Original Coordinates}

In the 2 dimensional case, $P_{m,n}(u,v) = P_{m,0}(u) P_{0,n}(v)$, the left and right edge polynomials of Figure \ref{uvlatt}.  The parameter $p$ only occurs in the function $P_{0,n}(v)$, so we re-label it as $P_{0,n}(p,v)$.  We then have
\begin{subequations}
\be\label{Pmnpur}
{\cal P}_{m,n}(p,u,r)= P_{m,0}(u) P_{0,n}\left(\frac{1}{2} (6p-1),\sqrt{v^2+w^2}\right)
\ee
Incorporating the orthogonal transformation (\ref{uvw}) and the gauge transformation, we have eigenfunctions
\be\label{phiqqq}
\phi_{m,n}(q_1,q_2,q_3) =  \left((q_1-q_2)(q_2-q_3)(q_3-q_1)\right)^{-p} e^{-\frac{1}{2} \omega (q_1^2+q_2^2+q_3^2)}\, {\cal P}_{m,n}(p,u,r),
\ee
\end{subequations}
with $u=\frac{1}{\sqrt{3}} (q_1+q_2+q_3),\, v^2+w^2 = \frac{1}{6}\left(3 (q_1-q_3)^2+(q_1-2 q_2+q_3)^2\right)$.  Combining the constant term of (\ref{Lhatcm3D}), with the Hermite eigenvalue of $-2m\omega$, we obtain $\lambda_{m,n}= (6p-2m-2n-3)\omega$.

\subsubsection*{Acknowledgments:}

I thank Oleg Chalykh for discussions regarding the Calogero-Moser system.


\begin{thebibliography}{10}

\bibitem{71-3}
F.~Calogero.
\newblock Solution of the one-dimensional n-body problems with quadratic and/or
  inversely quadratic pair potentials.
\newblock {\em J. Math. Phys.}, 12:419--36, 1971.

\bibitem{14-2}
M.~Cariglia.
\newblock Hidden symmetries of dynamics in classical and quantum physics.
\newblock {\em Rev.Mod.Phys.}, 86:1283--1333, 2014.

\bibitem{13-3}
M.~Feigin, O.~Lechtenfeld, and A.P. Polychronakos.
\newblock The quantum angular {Calogero-Moser} model.
\newblock {\em J. High Energ. Phys.}, 2013:162, 2013.

\bibitem{f05-1}
A.P. Fordy.
\newblock Symmetries, ladder operators and quantum integrable systems.
\newblock {\em Glasgow Mathematical Journal}, 47A:65--75, 2005.

\bibitem{f06-1}
A.P. Fordy.
\newblock Darboux related quantum integrable systems on a constant curvature
  surface.
\newblock {\em J.Geom.Phys.}, 56:1709--27, 2006.

\bibitem{f07-1}
A.P. Fordy.
\newblock Quantum super-integrable systems as exactly solvable models.
\newblock {\em SIGMA}, 3:025, 10 pages, 2007.

\bibitem{f18-3}
A.P. Fordy.
\newblock Classical and quantum super-integrability: From {Lissajous} figures
  to exact solvability.
\newblock {\em Atom. Nuclei}, 81:832 -- 42, 2018.
\newblock preprint arXiv:1711.10583 [nlin.SI].

\bibitem{f13-1}
A.P. Fordy and M.J. Scott.
\newblock Recursive procedures for {Krall-Sheffer} operators.
\newblock {\em J.Math.Phys}, 54:043516 (23 pages), 2013.

\bibitem{01-1}
J.~Harnad, L.~Vinet, O.~Yermolayeva, and A.~Zhedanov.
\newblock Two-dimensional {Krall-Sheffer} polynomials and integrable systems.
\newblock {\em J.Phys.A}, 34:10619--25, 2001.

\bibitem{40-2}
J.M. Jauch and E.L. Hill.
\newblock On the problem of degeneracy in quantum mechanics.
\newblock {\em Phys.Rev}, 57:641--5, 1940.

\bibitem{67-2}
H.L. Krall and I.M. Sheffer.
\newblock Orthogonal polynomials in two variables.
\newblock {\em Ann.Mat.Pura Appl. ser. 4}, 76:325--76, 1967.

\bibitem{96-5}
V.B. Kuznetsov.
\newblock Hidden symmetry of the quantum {Calogero-Moser} system.
\newblock {\em Physics Letters A}, 218:212 -- 22, 1996.

\bibitem{13-2}
W.~Miller Jr, S.~Post, and P.~Winternitz.
\newblock Classical and quantum superintegrability with applications.
\newblock {\em J.Phys.A}, 46:423001, 97 pages, 2013.

\bibitem{scott}
M.J. Scott.
\newblock {\em Classical and quantum integrable systems on manifolds with
  symmetry}.
\newblock PhD thesis, University of Leeds, 2010.

\bibitem{97-9}
J.F. van Diejen.
\newblock Confluent hypergeometric orthogonal polynomials related to the
  rational quantum {Calogero} system with harmonic confinement.
\newblock {\em Comm. Math. Phys.}, 188:467--497, 1997.

\bibitem{83-11}
S.~Wojciechowski.
\newblock Superintegrability of the {Calogero-Moser} system.
\newblock {\em Phys.Letts.A}, 95:279--81, 1983.

\end{thebibliography}

\end{document}